# Charge compensation by phase segregated $Sb_2Se_3$ phase in $Bi_{1.95}Sb_{0.05}Se_3$ topological insulator by laser fluence

Shivani Soni[1,2], Akhilesh Kumar[1,2], Edward Prabu Amaladass[1,2], Jegadeesan P[1], S. Amirthapandian[1,2], Kishore K. Madapu[1,2], Ramanathaswamy Pandian[1,2], Awadhesh Mani[1,2]

[1]Material Science Group, Indira Gandhi Centre For Atomic Research, Kalpakkam, 603102
[2]Homi Bhabha National Institute, Training School Complex, Anushaktinagar, Mumbai, 400094

E-mail: edward@igcar.gov.in

**Abstract:**

We report the charge compensation in topological insulator thin films due to the laser fluence-induced segregation of the $Sb_2Se_3$ phase. Sb doped $Bi_2Se_3$ films were deposited on commercial Si substrates coated with 300 nm of amorphous $SiO_2$ ($Si/SiO_2$) using the Pulsed Laser Deposition (PLD) technique at different laser fluence of 1.25 J-cm$^{-2}$, 1.87 J-cm$^{-2}$, 2.75 J-cm$^{-2}$, and 3.25 J-cm$^{-2}$. The grazing x-ray diffraction (GI-XRD) measurements revealed the growth of rhombohedral $Bi_2Se_3$ films on $SiO_2$ ($Si/SiO_2$) with additional peaks corresponding to the $Sb_2Se_3$ peaks. The phase fraction of the segregated $Sb_2Se_3$ systematically reduces as the laser fluence increases. The magnetoresistance measured at temperatures 3K - 20K was compared across different samples deposited at various laser fluences. The sample deposited at 1.25 J-cm$^{-2}$ exhibits a bulk-insulating nature with an order decrease in the carrier density. The weak antilocalization quantum correction to the magneto conductivity was analyzed using the Hikami–Larkin–Nagaoka Theory. As the laser fluence increases the deduced value of coefficient 'α' (number of conduction channels) was found to be 0.5, 1.5, 2.5, and 5.0, and the quantum phase coherence length '$l_\varphi$' ranges from 10 nm to 150 nm. The power factor ($\gamma$) of the temperature dependence of $l_\varphi$ ($l\varphi = T^{-\gamma}$) also varies as 0.37, 0.45, 0.69, and 0.92, respectively. The charge compensation due to the non-TI, p-type $Sb_2Se_3$ phase is believed to cause the reduction in the bulk carrier density (n), and enhanced surface state contribution.

**Introduction:**

Topological insulators exhibit a bulk band gap like conventional insulators but possess edges or surface conducting states within the band gap [1]. The topological surface states (TSS) of a 3D topological insulator $Bi_2Se_3$ are protected by time-reversal symmetry (TRS), making them robust against backscattering. $Bi_2Se_3$ has a simple electronic band structure with a single Dirac cone surface state and a small band gap of approximately 0.3 eV [2]. The discovery of surface states in 3D topological insulators with linear dispersion has opened up new possibilities for applications in spintronics, quantum computing[3], and also exhibits quantum effect like quantum anomalous Hall effect[4]. $Bi_2Se_3$ has a layered structure, wherein the quintuple layer (QL) is bound together weakly through van der Waals (VdW) interactions. Consequently, the selenium (Se) atoms terminating a QL are readily prone to escape from the crystal, leaving behind the Se vacancy defect ($V_{se}$), which acts as an electron donor. The high concentration of n-type charge carriers from these $V_{se}$ defects overshadows the material's intrinsic semiconducting properties, making the metallic characteristics of the surface states challenging to observe. Alloying is a well-studied method for reducing bulk electronic contributions in the materials. For example, doping $Bi_2Se_3$ with Te where $Bi_2Te_3$ is also a TI has shown a bulk insulating effect [5, 6]. $Bi_2Te_3$ an isostructural material, can be synthesized as p-type and $Bi_2Se_3$ is typically n-type because of Se vacancy. So doping of $Bi_2Te_3$ in $Bi_2Se_3$ also helps in charge compensation. In addition, Se escape is reduced in $Bi_2SeTe_2$ ternary compounds due to the modified atomic arrangement in the crystal [5] thereby reducing the overall number of carriers. Similarly, doping Sb at the Bi site is also reported to reduce the carrier density [7]. Quaternary TI composed of Bi, Se, Sb, and Te has been reported to possess better TI properties with bulk insulating nature [5]. Similarly, Indium doping in $Bi_2Se_3$ ($Bi_{1-x}In_x)_2Se_3$ reduces carrier density by introducing acceptor states, shifting the Fermi level towards the intrinsic regime, and modifying defect-assisted charge compensation mechanisms [8]. The study of various defects enables the accurate identification of the nature of native defects and their relative concentrations, including selenium vacancies ($V_{Se}$), bismuth antisites ($Bi_{Se}$), and selenium interstitials ($Se_i$)[9]. It is established that the dominant $V_{Se}$ defect is positioned in the middle of the quintuple layer, rather than at the energetically favorable internal van der Waals (VdW) surfaces. Combined with the presence of excess selenium atoms between the layers, this highlights the role of kinetics over thermodynamics in defect formation within $Bi_2Se_3$, and potentially in other chalcogenide-based topological insulators..

Thin films of $Bi_2Se_3$ offer more control over the contribution of bulk carriers to the conductance, making them attractive for studying the surface states. Studying the system in thin film form is essential to increase the surface-to-bulk ratio and for large-scale fabrication of topological insulators (TIs) suitable for device. The growth of oriented $Bi_2Se_3$ thin films has been documented using several techniques, each with distinct advantages and disadvantages. Notable methods include Molecular Beam Epitaxy (MBE) [10-12], Magnetron Sputtering [13, 14], and Pulsed Laser Deposition (PLD) [15-18]. Here we have studied the effect of the laser fluences on the transport properties of the PLD deposited $Bi_{1.95}Sb_{0.05}Se_3$ thin films.

**Experimental Details:**

The PLD target $Bi_{1.95}Sb_{0.05}Se_3$ (BSS) was prepared using a single crystal grown by the self-flux method. The crystals were ground and pelletized to form PLD targets with a diameter of 10 mm[19]. The BSS thin film was deposited onto a $Si/SiO_2$ substrate, coated with a 300 nm layer of amorphous $SiO_2$, utilizing the PLD technique. Before deposition, the substrate was cleaned by boiling in deionized (DI) water for 3 minutes, followed by ultrasonic cleaning in DI water, acetone, and isopropyl alcohol. Ablation of the target material was performed using a KrF excimer laser ($\lambda$ = 248 nm) at a repetition rate of 5 Hz. The thin films were deposited in an Argon atmosphere at a pressure of 0.5 mbar, with the substrate temperature maintained at 350°C and a fixed target-to-substrate distance of 3 cm. The laser fluence was systematically varied as 1.25 J-cm$^{-2}$, 1.87 J-cm$^{-2}$, 2.75 J-cm$^{-2}$, and 3.25 J-cm$^{-2}$. The BSS thin films were deposited under the defined conditions to investigate the influence of PLD parameters on their surface morphology and transport properties. The films were analyzed using a Carl Zeiss Crossbeam 340 scanning electron microscope (SEM), capturing images in in-lens duo mode at accelerating voltages of 3–5 kV. Glancing angle X-ray diffraction (GI-XRD), studies were conducted to evaluate the structural properties of the films using a Cu-K$\alpha$ source with a wavelength of 1.54 A° at a glancing angle of 0.5° in a Stoe diffractometer. Raman measurements were carried out at room temperature (RT) with the excitation of a 532 nm laser (InVia, Renishaw). Electrical transport measurements were performed using a cryo-free magnetoresistance setup from Cryogenic UK, in perpendicular magnetic fields in the range of +5 T to -5 T and at various temperatures from 3 K to 300 K. Room temperature scanning tunneling microscopy (STM) (nanoREV, Quazar Technologies Pvt. Ltd., New Delhi India) is utilized to investigate the electronic properties of the sample and the experiment is conducted at the vacuum of $10^{-2}$ mbar.

**Result and Discussion:**

Figure. 1 shows the grazing incidence X-ray diffraction (GI-XRD) analysis of antimony-doped bismuth selenide ($Bi_{1.95}Sb_{0.05}Se_3$) thin films deposited at laser fluence of 1.25 J-cm$^{-2}$, 1.87 J-cm$^{-2}$, 2.75 J-cm$^{-2}$, and 3.25 J-cm$^{-2}$. All diffraction peaks have been indexed to $Bi_2Se_3$, referencing the ICDD file no. 00-033-0214. Which crystallizes in a rhombohedral structure (R-3m)  with quintuple layers arranged in a $Se_1$-Bi-$Se_2$-Bi-$Se_1$ sequence and is a well-known topological insulator[2]. We observed an additional peak at a higher angle of 64.2 – 64. 4 as shown in **Error! Reference source not found.**.**Error! Reference source not found.**

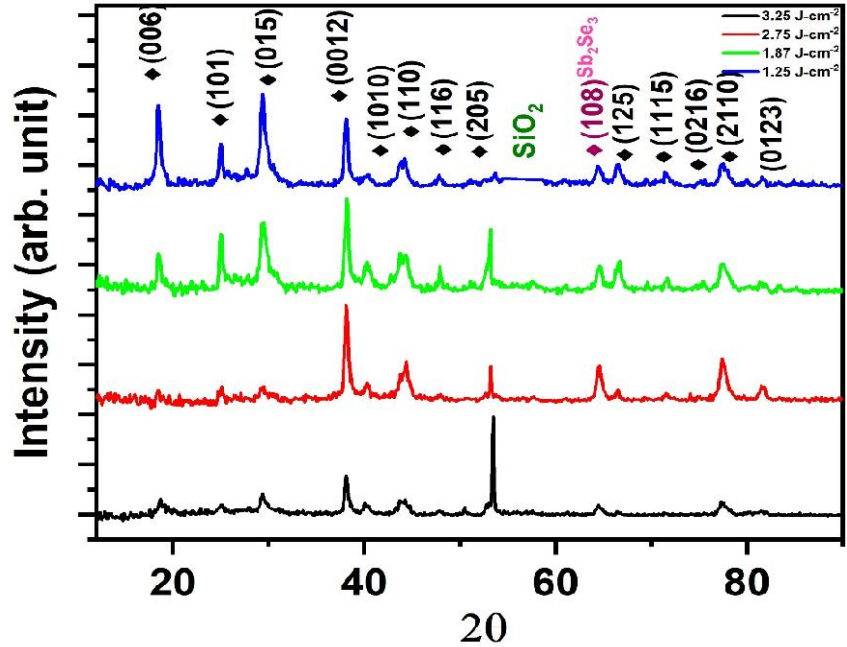

This higher-angle peak corresponds to $Sb_2Se_3$, indexed to ICDD file no. 01-075-1462.  $Sb_2Se_3$ crystallizes in an entirely different orthorhombic structure(space group Pnma)[20] and lacks topological insulating properties[21]. The crystal lattice of $Sb_2Se_3$ is composed of one-dimensional (1D) chains of $(Sb-Se)_n$ aligned along the c-axis, forming linear structures[22]. These chains are held together by weak van der Waals forces, resulting in anisotropic properties, where the material exhibits different behavior along different crystallographic axes. The phase fraction of $Sb_2Se_3$ was estimated by fitting the XRD peaks with the Gaussian function and the area under the curve. The resulting phase fractions are 2.86%, 3.53%, 1.61%, and 0.5%, for films deposited at 1.25 J-cm$^{-2}$, 1.87 J-cm$^{-2}$, 2.75 J-cm$^{-2}$, and 3.25 J-cm$^{-2}$ respectively.The phase fraction of the $Sb_2Se_3$ phase is found to be higher in the films deposited with lesser laser fluence.

In order to study the surface morphologies of these films, SEM images were recorded. *Figure 2* shows the SEM images and the inset shows the log-normal distribution of the estimated grain size using Image J software. As laser fluence increases grain size increases from 114 nm to 129 nm. The increase in grain size and the observation of more faceted grains at higher laser suggest better crystalline samples at higher laser fluence. However, the phase-segregated $Sb_2Se_3$ species could not be differentiated due to similar atomic numbers. Cross-sectional SEM indicates film thickness variations of 193 nm, 194 nm, 215 nm, and 218 nm as shown in the bottom panel of *Figure 2*.

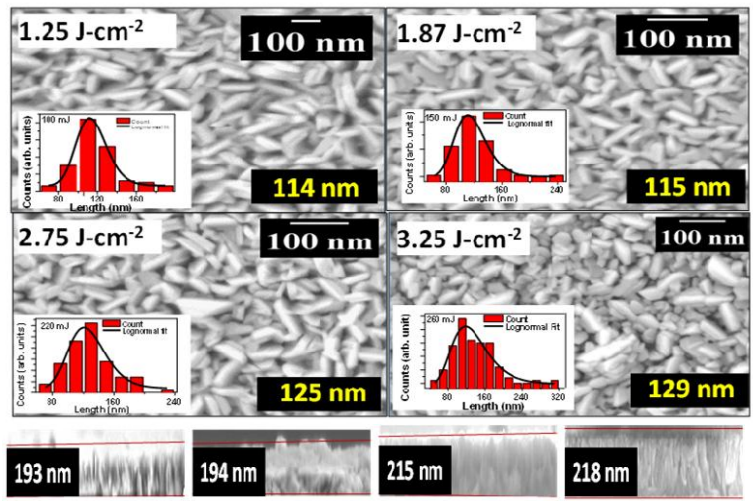

Figure 1: *The top panel shows SEM images of BSS thin films deposited at different laser fluence. The insets show the grain size distribution fitted to a lognormal function. The bottom panel shows the thickness of the film (cross-sectional SEM)*

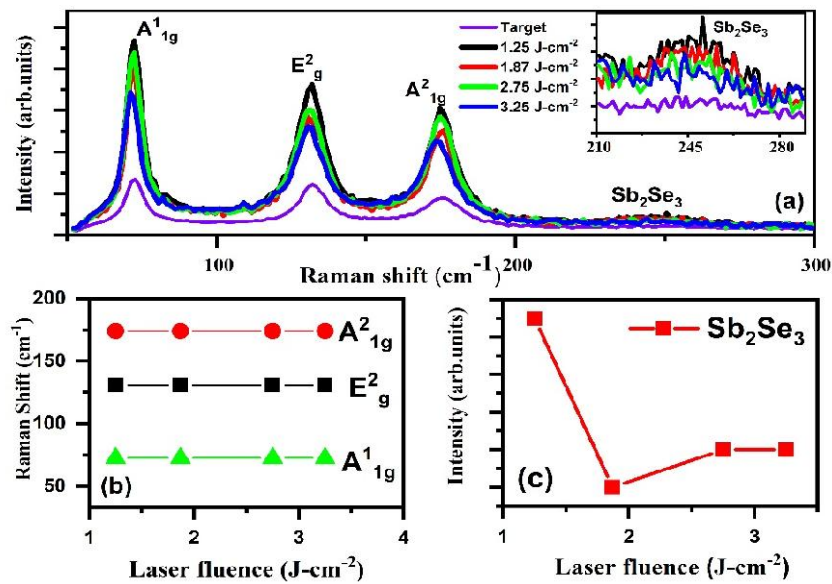

Figure 3: *(a) Characteristic Raman spectra of BSS films deposited at different laser fluences: 1.25 J-cm$^{-2}$, 1.87 J-cm$^{-2}$, 2.75 J-cm$^{-2}$, and 3.25 J-cm$^{-2}$ with the inset highlighting Raman modes corresponding to the Sb$_2$Se$_3$ phase. (b) Raman shift as a function of laser fluence for different Bi$_2$Se$_3$ modes. (c) Decrease in Sb$_2$Se$_3$ phase intensity with increasing laser fluence.*

To get more insights, Raman spectra were recorded at room temperature. The peaks were analyzed by fitting with the Lorentzian function. The spectra in Figure 3(a) reveal three distinct Raman peaks in the low wavenumber region, corresponding to the vibrational modes of $Bi_2Se_3$: $A^1_{1g}$, $E^2_g$, and $A^2_{1g}$ and the inset shows the Raman modes corresponding to $Sb_2Se_3$. Three Raman modes with $E_g$ and $A_{1g}$ symmetries are observed. The $E_g$ mode is the in-plane vibration of Bi-$Se_1$ pairs, and the two $A_{1g}$ modes are due to out-of-plane vibrations of Bi-$Se_1$ pairs [23, 24]. As laser fluence increases, intensity decreases as shown in Figure (a). An increase in laser fluence correlates with a rise in the full width at half maximum (FWHM) of the peaks, resulting in peak broadening, which can be due to the increased defects attributed to enhanced phonon-phonon interactions. Interestingly, a prominent peak corresponding to the $Sb_2Se_3$ phase exhibits maximum intensity at lower laser fluence.

Table 1: Room temperature Raman data of $Bi_2Se_3$ thin film on $Si/SiO_2$ substrate

| Sample at a different laser fluence (J-cm$^{-2}$) | $A^1_{1g}$ | | $E^2_g$ | | $A^2_{1g}$ | |
|---|---|---|---|---|---|---|
| | Peak position (cm$^{-1}$) | FWHM (cm$^{-1}$) | Peak position (cm$^{-1}$) | FWHM (cm$^{-1}$) | Peak position (cm$^{-1}$) | FWHM (cm$^{-1}$) |
| 1.25 | 72.31 | 6.51 | 131.30 | 13.14 | 175.05 | 12.81 |
| 1.87 | 72.34 | 6.57 | 131.37 | 13.66 | 175.04 | 12.13 |
| 2.75 | 72.28 | 6.73 | 130.94 | 14.38 | 175.07 | 12.76 |
| 3.25 | 71.90 | 7.73 | 130.57 | 16.77 | 173.68 | 14.53 |

However, as laser fluence increases, the intensity of this peak gradually decreases and vanishes because, at elevated laser fluence, the increased kinetic energy of the atoms facilitates the diffusion of Sb atoms into the $Bi_2Se_3$ lattice. In addition we also believe that the enhanced mobility of abalated species lowers the energy barriers associated with the substitution of Sb for Bi within the $Bi_2Se_3$ structure. When energy falls below a certain threshold known as the solubility limit, the alloy may no longer accommodate the excess Sb within the $Bi_2Se_3$ structure, resulting in the precipitation or segregation of Sb into a secondary phase, most commonly $Sb_2Se_3$. $Sb_2Se_3$ typically forms at intermediate to high temperatures, ranging from 400 to 600°C [25] and remains stable over a broad temperature range. $Sb_2Se_3$ is a stable binary compound that forms along the Sb-Se edge of the ternary Bi-Se-Sb phase diagram, with a 2:3 molar ratio of antimony to selenium[25, 26]. Thermodynamically, higher energies can promote the stability of the solid solution phase, allowing a greater concentration of Sb atoms to dissolve in the $Bi_2Se_3$ matrix without leading to the formation of a separate phase such as $Sb_2Se_3$. In the $Bi_{1.95}Sb_{0.05}Se_3$ alloy, the introduction of 5% Sb is intended to replace Bi in the $Bi_2Se_3$ lattice for lesser Se escape. However, the solubility of Sb in $Bi_2Se_3$ is relatively low; therefore, when this limit is exceeded, any excess Sb is likely to segregate and form a distinct $Sb_2Se_3$ phase[25]. In the Sb-Se binary phase diagram, $Sb_2Se_3$ forms at compositions with approximately 40% antimony and 60% selenium[25]. In the ternary Bi-Se-Sb system, $Sb_2Se_3$ forms a distinct phase with a narrow, well-defined compositional range. When Sb and Se are mixed in the correct proportions, $Sb_2Se_3$ remains stable with minimal solubility in excess Sb or Se. In the presence of Bi, $Sb_2Se_3$ can still form at low Bi concentrations; however, higher Bi levels lead to phase competition, where both $Bi_2Se_3$ and $Sb_2Se_3$ phases coexist[25]. The formation of a distinct $Sb_2Se_3$ phase is thermodynamically driven. The alloy system minimizes its free energy by allowing excess Sb to segregate into its stable form, $Sb_2Se_3$, rather than attempting to dissolve entirely within the $Bi_2Se_3$ structure. In the case of the $Sb_2Se_3$ phase formation, the excess Sb segregates to minimize the overall free energy of the system, resulting in a more stable arrangement that is less likely to change under equilibrium conditions.

Notably, the graph indicates no shift in the peak positions with laser fluence as shown in Figure 3 (b). The peak positions and their corresponding full-width half maxima (FWHM) values for the $A^1_{1g}$, $E^2_g$, and $A^2_{1g}$ modes provide insight into the structural and electronic properties of the material [27], which are listed in

Table 1. Remarkably, the FWHM values for these significant vibrational modes across all four samples are slightly elevated compared to those reported in the literature. As the laser fluence increases, the intensity of the $Sb_2Se_3$ peak decreases. The phase fraction of $Sb_2Se_3$ is quantitatively determined from the Raman modes using Equation 1, yielding values of 5.7 %, 4.8 %, 5 %, and 5% with decreasing laser fluence, as shown in Figure 3(c). To assess the extent of segregation observed in the Raman spectra and its potential impact on elemental composition, Energy Dispersive Spectroscopy (EDS) was performed. The EDS results, as shown in Table 2, did not reveal significant variations in elemental

$$\% Sb_2Se_3 = \frac{I(A^1_{1g}) + I(E^2_g)}{I(A^1_{1g}) + I(E^2_g) + I(Sb_2Se_3)} \quad \text{———— Equation-1}$$

composition. This shows that the segregated phase fraction of $Sb_2Se_3$ is below the detection resolution of the EDS measurement or we get the averaged values corresponding to both the phases.

Table 2: Atomic Percentage of Bi, Se, and Sb obtained from EDS analysis:

| Growth Condition Sub temp - 350°C, Ar press -0.5 mbar | Atomic Percentage of Bi (%) | Atomic Percentage of Se (%) | Atomic Percentage of Sb (%) |
|---|---|---|---|
| 1.25 J-cm$^{-2}$ | 40.5 | 59 | 0.5 |
| 1.87 J-cm$^{-2}$ | 40.5 | 59 | 0.5 |
| 2.75 J-cm$^{-2}$ | 41 | 58 | 1 |
| 3.25 J-cm$^{-2}$ | 41.5 | 58 | 0.5 |

To further investigate the role of the phase segregated $Sb_2Se_3$ phase, the temperature and magnetic field dependence of the longitudinal and transverse resistance were studied. Figure shows the resistivity vs temperature plot of the BSS sample deposited at different laser fluence. The sample prepared at 1.25 J-cm$^{-2}$ exhibits an insulating behavior; in contrast, samples prepared at higher laser fluence that show a metallic trend.

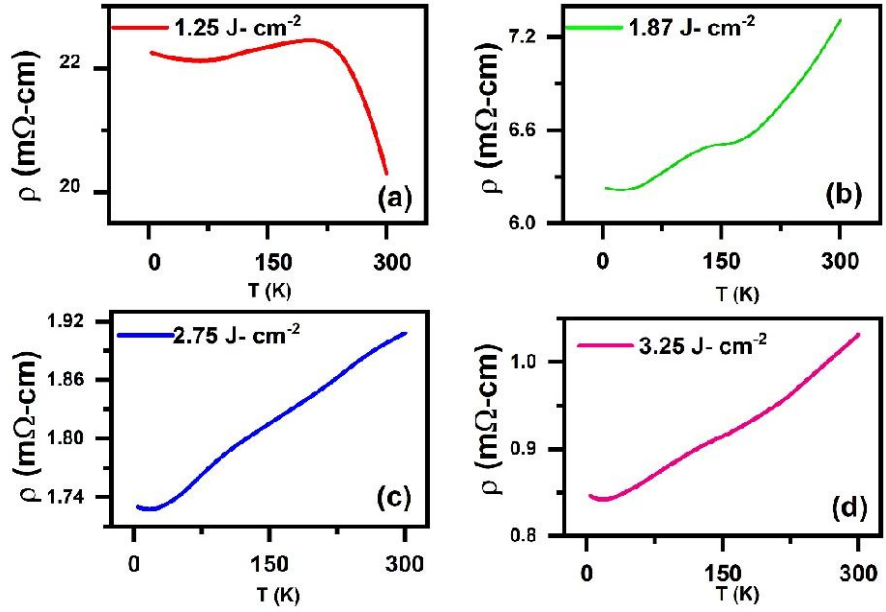

*Figure 4: Resistivity as a function of temperature for (BSS) thin films grown at varying laser fluence: (a) 1.25 J-cm⁻² (b) 1.87 J-cm⁻² (c) 2.75 J-cm⁻², and (d) 3.25 J-cm⁻². The measurements were performed over a temperature range from low temperature 4.2 K to room temperature 300 K.*

Figure shows the Hall data measured at different temperatures. Hall measurements were conducted within a magnetic field range of ±5 tesla to determine the charge carrier concentration. The negative slope indicates the presence of n-type charge carriers, typically associated with selenium (Se) vacancies. The carrier density was determined using:

$$\mathbf{n} = \frac{1}{R_H e}$$

where n represents the carrier density, e denotes the electronic charge, and $R_H$ refers to the Hall coefficient. As laser fluence increases, the carrier density also shows an increasing trend in Figure 5 (e). The calculated carrier density showed a one-order change at lower laser fluence from $10^{19}$ cm⁻³ compared to higher laser fluence ranging to $10^{20}$ cm⁻³. This observation highlights the charge compensation effect of bulk charge carriers. The $Sb_2Se_3$ phase is non-topological and it is a p-type with holes as the majority of carriers [28]. As the phase fraction of p-type $Sb_2Se_3$ is higher in the sample prepared at lower laser fluence the bulk carriers show a one-order decrease due to electron and hole compensation. This is believed to suppress the bulk contribution and enhance the metallic surface state conductance. The Hall mobility was deduced using the formula:

$$\mu = \frac{1}{\rho n e}$$

where ρ represents the resistivity, n denotes the carrier density, and e is the charge of an electron. As seen in Figure 5 (f) mobility initially rises however, at higher laser fluence, it begins to decrease once again.

$Sb_2Se_3$ is a non-topological phase with a band gap of 1. 2 ev coexisting with the TI BSS matrix of 0.3 ev band gap with a rhombohedral phase.   However, the fact that there is a one-order change in the carrier density underscores that there is a charge compensation when the concentration of the $Sb_2Se_3$ phase is comparably higher.

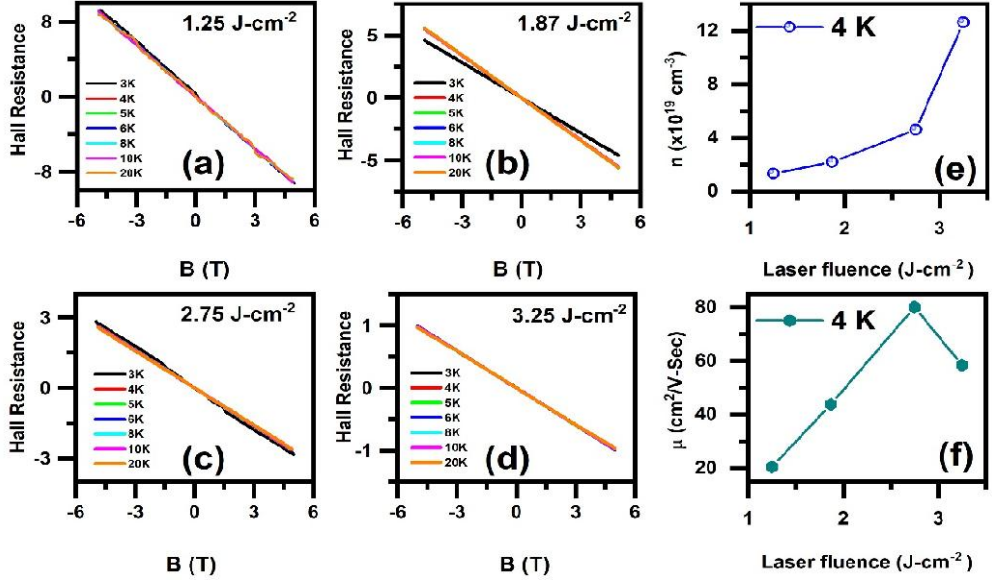

Figure 5: Hall data of four representative samples at different laser fluence: *(a) 1.25 J-cm⁻², (b) 1.87 J-cm⁻², (c) 2.75 J-cm⁻², and (d) 3.25 J-cm⁻²*. Panels (e) and (f) present a comparison of carrier density and mobility, respectively, as a function of laser fluence at low temperatures (4 K).

The weak antilocalization (WAL) phenomenon in the topological insulator is a viable indirect tool to access the surface state properties. Magnetoresistance measurements were performed to identify and study WAL features in the samples . Temperature-dependent MR data in  Figure 6 (a-d) measured in the perpendicular magnetic field ranging from -5 T to +5 T at temperatures from 3 K to 10 K shows a sharp cusp due to weak antilocalization. The WAL features were fitted with the Hikami- Larkin- Nagaoka equation [29], which is as given below in the field ranging from -1 T to +1 T.

$$\Delta G\ (B) = -\alpha\ \frac{e^2}{2\pi^2 \hbar}\left[\psi\left(\left(\frac{\hbar}{4el_\phi^2 B}\right) + \frac{1}{2}\right) - ln\left(\frac{\hbar}{4el_\phi^2 B}\right)\right]$$

Here, $\Delta G\ (B)$  is the magnetoconductance at a given magnetic field B. $\Psi$ is the digamma function, $\hbar$ the reduced Planck's constant, e the electronic charge, and $l_\phi$ and $\alpha$ the parameters obtained through the fitting. Where,  $l_\phi$ is the quantum coherence length, and $\alpha$ characterizes the number of independent channels. The magneto-conductance term ΔG(B) which is a result of the cyclotron motion of charge carriers, is mathematically defined as the difference between the zero-field conductance G(0) and the conductance under a magnetic field G(B).

$$\Delta G(B) = G(B) - G(0)$$

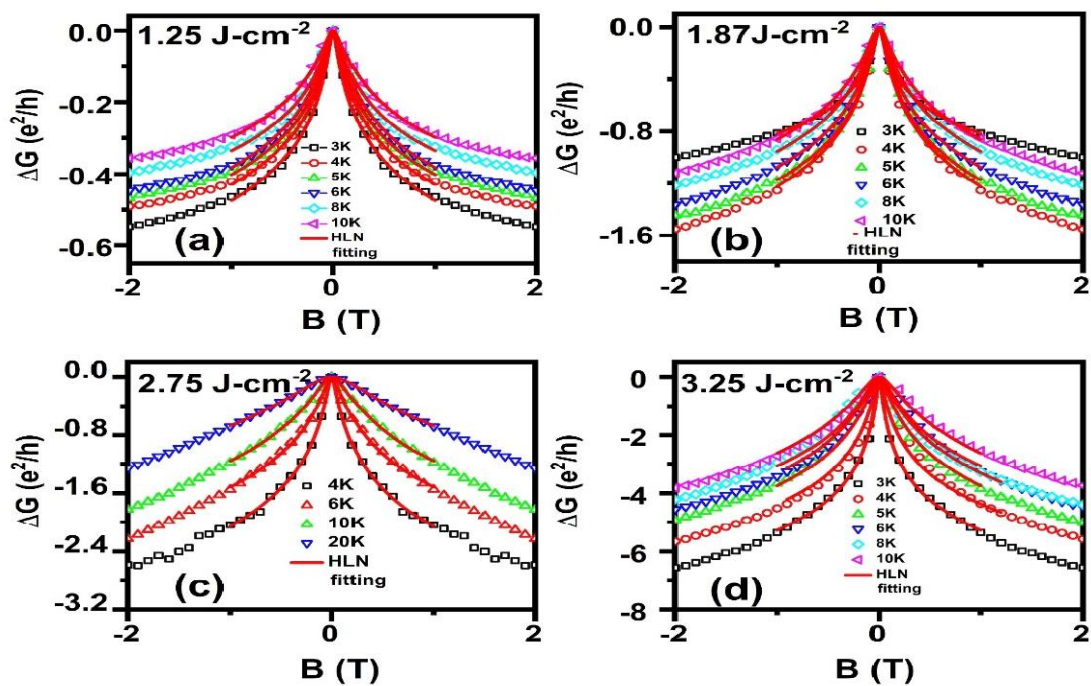

*Figure 6*: Magnetoconductance as a function of the magnetic field for BSS thin films grown at different laser fluence and data is fitted using the Hikami-Larkin-Nagaoka (HLN) equation, illustrated in plots (a), (b), (c), and (d).

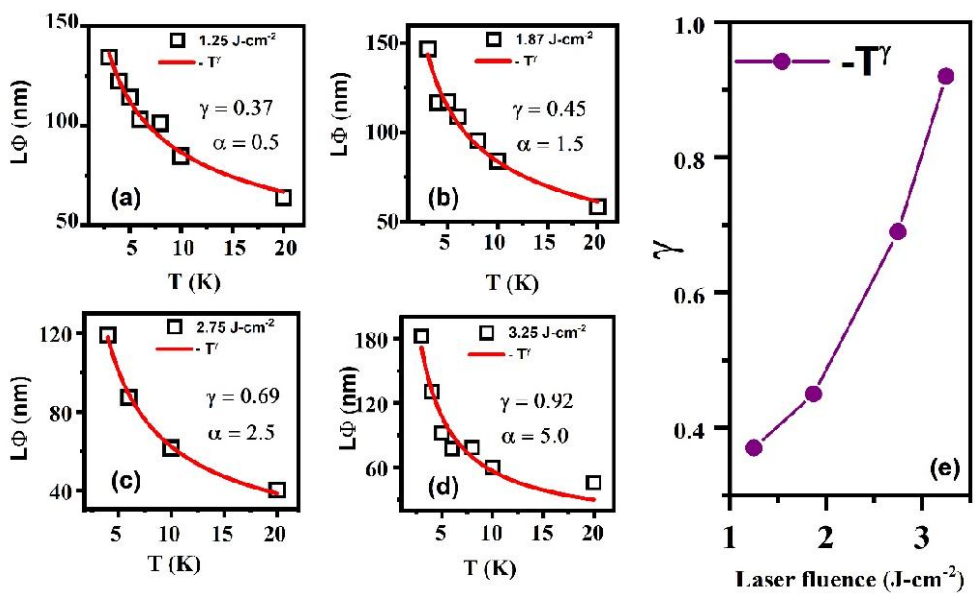

*Figure 7: Phase coherence length extracted from weak antilocalization (WAL) fits as a function of temperature for samples grown at different laser fluence: (a) 1.25 J-cm⁻², (b) 1.87 J-cm⁻², (c) 2.75 J-cm⁻², and (d) 3.25 J-cm⁻². The gamma exponent (γ), derived from the temperature dependence of the phase coherence length $L_\phi \propto T^{-\gamma}$ is presented as a function of laser fluence in plot (e).*

The ΔG (B) of BSS films are plotted against the magnetic field (B) in Figure 6 (a-d). The temperature dependence of $l_\varphi$ obeys a power law ($l_\varphi = T^{-\gamma}$), where γ the exponent value should be 0.5 for an ideal topological insulator. It means there is only electron-electron interaction as a prime dephasing mechanism in the system. This is used as the yardstick to access the surface states in the system. The $l_\varphi(T)$, for all the four samples, are shown in Figure 7 (a-d). It was observed that the exponent γ differed, with values of 0.37, 0.45, 0.69, and 0.92. Notably, at 1.25 J-cm⁻² and 1.87 J-cm⁻² where (γ = 0.37, 0.45) is close to 0.5, it indicates that electron-electron interactions are the predominant cause of dephasing in the system. As the laser fluence increases, this power-law exponent also rises, suggesting that the higher-order dephasing terms like electron-phonon interactions become more significant, indicating increased bulk contribution. In addition, the fitting parameter α also indicates the number of coherent channels in the transport. For a single Dirac surface state, it is 0.5. The samples prepared at the lower laser fluence showed the α values close to 0.5 pointing to the fact the contribution of coherent surface states in the WAL phenomenon. To further investigate the fluence-dependent modifications in electronic and structural properties, we conducted scanning tunneling microscopy studies (STM). The STM morphology in Figure 8 (a-d) provide nanoscale insights into morphological variations, including grain connectivity and defect density, which play a crucial role in governing charge transport. These observations establish a direct correlation between the magnetotransport characteristics and the structural evolution of the films, emphasizing the role of laser fluence in modulating the interplay between bulk and surface states.

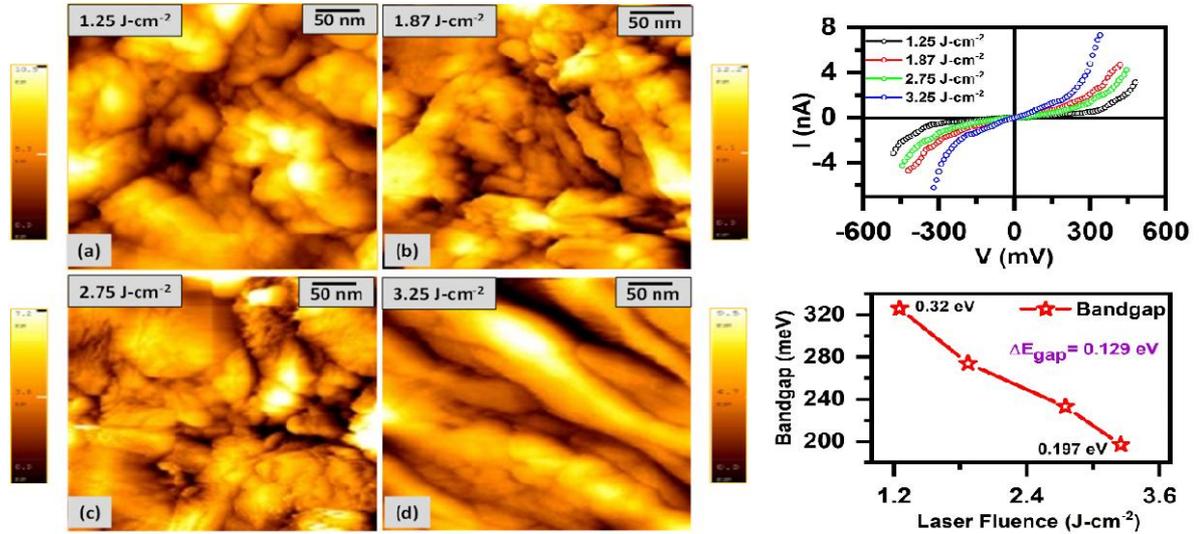

*Figure 8: Scanning tunneling Microscope images (a-d) of sample deposited at different laser fluence, The right panel shows the corresponding I-V characteristics (e) and bandgap variation with laser fluence (f).*

At lower laser fluences 1.25 J-cm$^{-2}$ and 1.87 J-cm$^{-2}$, the grains are small and well-separated, while at higher fluences 2.75 J-cm$^{-2}$ and 3.25 J-cm$^{-2}$, the grains become more elongated and interconnected. The I-V characteristics measured at different laser fluence indicate variation in electronic transport behavior Figure 8(e). At lower laser fluence, the current is lower for a given voltage, suggesting a higher bandgap and insulating behavior. As the laser fluence increases, the current increases, indicating enhanced conductivity, which correlates with the observed bandgap reduction. The bandgap analysis (f) shows that at lower laser fluence, the bandgap is approximately 0.32 eV, while at higher fluence, it decreases to 0.197 eV, corresponding to a reduction of 0.129 eV. This suggests that at lower laser fluence, Sb$_2$Se$_3$ segregation is more prominent, leading to compensation of the Sb$_2$Se$_3$ acceptor states with the donor V$_{Se}$ state. At higher laser fluence, the increased incorporation of Sb into the Bi$_2$Se$_3$ lattice alters the electronic band structure thereby reducing the bandgap. The average surface roughness of the film, as determined by STM, is 6 nm.

The role of charge compensation due to the phase-segregated p-type Sb$_2$Se$_3$ in the n-type BSS matrix can be understood as follows. A schematic representing the shift in Fermi energy level in the BSS system is illustrated in **Error! Reference source not found.**9.

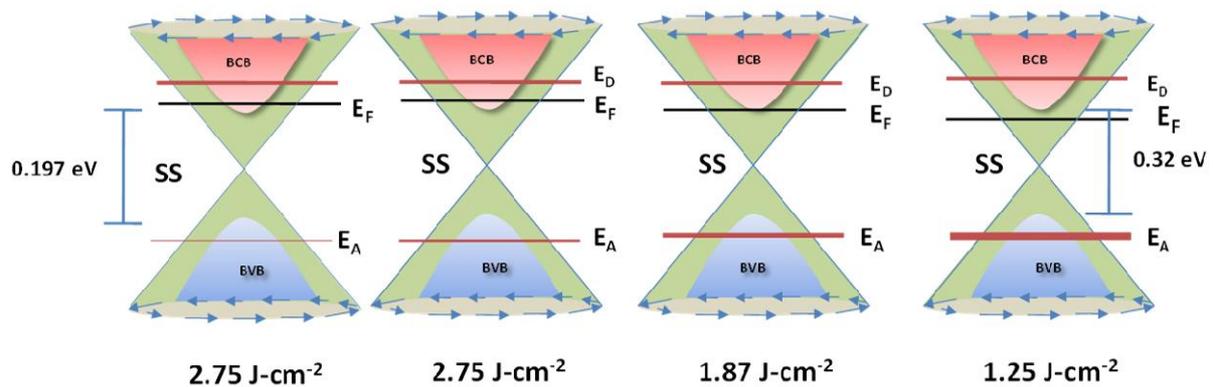

*Figure 9: Shift in Fermi energy level in the BSS system due to charge compensation arising from phase segregation of Sb$_2$Se$_3$*

Here, BCB refers to the Bulk Conduction Band, and BVB refers to the Bulk Valence Band. E$_F$ represents the Fermi level, E$_D$ is the donor level, and E$_A$ is the acceptor level. The donor level is associated with selenium vacancies (V$_{Se}$) or interstitial selenium (Se$_i$), while the acceptor level corresponds to Sb$_2$Se$_3$. The segregation of the Sb$_2$Se$_3$ phase in the system leads to the formation of acceptor levels and thereby compensating the excess carrier from the bulk. As a result, the Fermi energy shifts downwards tending towards the quasi-intrinsic regime, and increases the contribution of the surface state in the conduction.

The role of temperature and the pressure during deposition on the phase separation phenomenon and its impact on the transport properties is believed to shed more insight into this system which will be carried out in the future.

**Conclusion:** Thin films of Bi$_{1.95}$Sb$_{0.05}$Se$_3$ were grown using a pulsed laser deposition (PLD) system at varying laser fluence. A systematic evolution of structural, electronic, and magnetotransport properties was observed with changes in laser fluence. The XRD and Raman analysis showed the existence of an additional $Sb_2Se_3$ phase in the sample. Its phase fraction is found to be higher in the samples deposited at lower laser fluence and it is systematically reduced as the laser fluence increases. The existence of this p-type phase in the n-type TI matrix is believed to cause charge compensation. As a result, the sample prepared at the lower fluence showed an insulating nature as compared to the other samples that were metallic. Carrier density also decreased by one order of magnitude in the sample with a higher phase fraction of $Sb_2Se_3$. HLN analysis of the WAL phenomenon showed that the power law exponent and the alpha are in line with the values of a good TI system.

**Acknowledgment:** One of the authors (Shivani Soni) would like to acknowledge the Department of Atomic Energy, India, for providing experimental facilities. We thank UGC-DAE CSR, Kalpakkam node, for providing access to magnetic and magnetotransport measurement systems.